\documentclass[11pt]{article}
\usepackage{epsfig}

\catcode`@=11
\catcode`@=12

\newcommand{\be}{\begin{equation}}
\newcommand{\ee}{\end{equation}}
\newcommand{\bref}[1]{(\ref{#1})}
\def\bl{\biggl(}
\def\br{\biggr)}
\def\ie{{\it i.e. }}
\def\dphi{\phi^{\dagger}}
\def\t{\theta}
\def\kvec{\vec{k}}
\def\pvec{\vec{p}}
\def\opvec{\vec{P}}
\def\del3{\delta^{(3)}}
\def\tp{\tilde{P}}
\def\otp{\tilde{p}}

\def\undos{{1 \over 2}}
\def\pa{\partial}

\def\CL{{\cal L}}
\def\CT{\Omega}
\def\lam{\lambda}
\def\e{\epsilon}

\begin{document}


\begin{titlepage}
\vfill
\begin{flushright}
UB-ECM-PF-01/02\\

\end{flushright}
\vfill

\begin{center}
\baselineskip=16pt
{\Large\bf  
A Note on Unitarity of Non-Relativistic 
Non-Commutative Theories}
\vskip 0.3cm
{\large {\sl }}
\vskip 10.mm
{\bf ~Toni Mateos$^1$, Alex Moreno$^2$}\\
\vskip 1cm
{\small
Departament ECM, Facultat de F\'{\i}sica\\
Universitat de Barcelona and Institut de F\'{\i}sica d'Altes
Energies,\\
Diagonal 647, E-08028 Barcelona, Spain}\\ 
\vskip.4cm
\end{center}
\vfill
\par

\begin{abstract}

We analyze the unitarity of a 
non-relativistic non-commutative scalar field theory. 
We show that electric
backgrounds spoil unitarity while magnetic ones
do not. Furthermore, unlike its relativistic
counterparts, unitarity can not be restored (at least
at the level of one-to-one scattering amplitude)
by adding new states to the theory. This is a signal that the
model cannot be embedded in a natural way in string theory.

\end{abstract}

\vfill{
 \hrule width 5.cm
\vskip 2.mm
{\small
\noindent $^1$ E-mail: tonim@ecm.ub.es \\
\noindent $^2$ E-mail: moreno@ecm.ub.es
}}

\end{titlepage}

\section{Introduction}

\indent

	It has been proved \cite{jaume1,noncausal} 
that non-commutative ($NC$) field theories
are non-unitary and acausal when the non-commutativity
involves the time coordinate. 
The non-locality in time requires a more
subtle hamiltonian formalism \cite {gkl,toni}
where one needs to add an extra time-like
coordinate. Except for several special cases
of light-like non-commutativity \cite{jaime}, 
this non-locality in time is responsible for spoiling
unitarity.

	From a string theoretical point of view,
these problems seem reasonably well understood. Magnetic
$B_{\mu\nu}$ backgrounds of string theory admit 
a decoupling limit of the massive modes \cite{m1,m2,m3}
and lead to magnetic non-commutative field theories 
in the worldvolume of the D-branes.
The unitarity of string theory is then inherited
by the effective field theory. 

On the other hand, electric backgrounds
of string theory do not admit such a decoupling limit
\cite{l1,l2,l3}.
The lack of unitary of the effective
field theory in the brane is just telling us
that we should have taken into
account all the massive string modes. In particular, it 
is found in \cite{luis} that adding
new stringy states (the so-called
$\chi$-particles) to a NC $\phi^4$ theory, 
one could understand and solve the unitarity 
failure at the level of one-to-one scattering
amplitude.

Our work was doubly motivated. First of all, we
wanted to check if these unitarity problems
are present in non-relativistic theories too,
where the treatment of space and time is completely
different. In this paper, we study
a non-relativistic $NC$
$\phi^4$ theory in $(2+1)$ dimensions, which
can be nicely viewed as the realization of the
Galileo group with two central extensions (the mass
and the non-commutativity parameter $\t$)\cite{luki}.  
We found an affirmative answer: 
only electric backgrounds break unitarity.
This is the subject of chapter two.

Secondly, we wanted to understand whether
unitarity could be restored by adding some new states.
In such case, they would probably come from
a non-decoupled non-relativistic
theory of open strings, like that in \cite{ncos,guijosa}. 
We found that such a procedure is not possible 
in our non-relativistic model, even for the 
simplest case of one-to-one scattering.
Better understanding is therefore still required.
We address this problem in section three.

\section{Four Points Function and Unitarity}

\indent 

To set up our framework, let us start with a non-relativistic
$NC$ scalar field theory in $D=2+1$ with quartic interactions
and Lagrange density  \footnote{The perturbative properties of 
this model in the magnetic case have been studied in \cite{esperanza}
and exact results can be found in \cite{bak}.
} $^,$ \footnote{
It can be seen that having taken the other possible
ordering of the vertex, \ie $\dphi*\phi*\dphi*\phi$, 
would have lead to exactly the same unitarity problems.} 

\be
\CL_{nr}=\dphi \left(i\pa_t + {\vec{\nabla}^2 \over 2} \right)
\phi - {\lambda \over 4} \, \dphi*\dphi*\phi*\phi.
\label{lnr}
\ee

Following the idea of \cite{jaume1}, we will study
the unitarity of the theory by checking whether the
Optical Theorem is fulfilled
at the level of two particles scattering.
It should happen that
\be  \epsfig {file=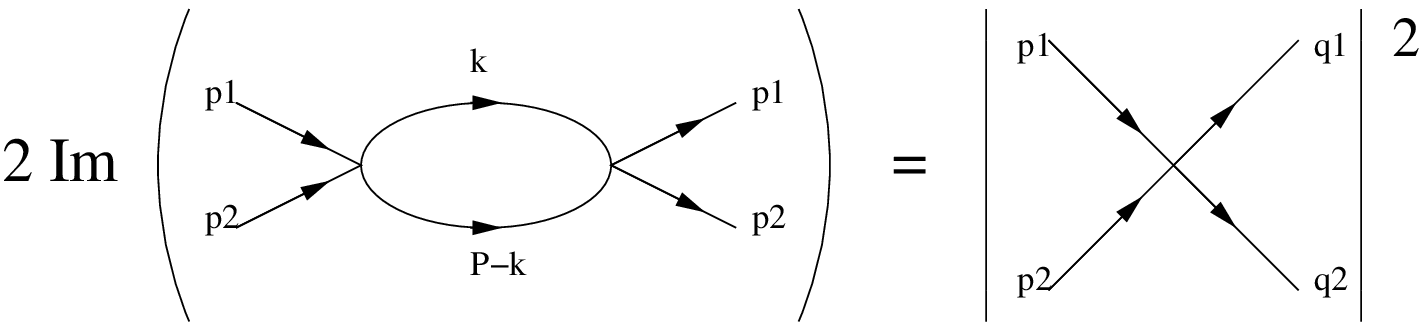, height=2cm, width=9cm} \ee
The left hand side ($LHS$) and the right hand side ($RHS$)
can be written as 
\be
LHS \equiv 2 Im \bl
-i \frac{\lambda^2}{2} \, cos^2 {\otp_1  p_2 \over2} \,
\int {d^2k dk^0 \over (2\pi)^3}\,
{cos^2 { \tp  k \over 2} \over
[k^0 - \frac{\kvec^2}{2} +i\e]
 [p^0-k^0 - \frac{(\pvec-\kvec)^2}{2} +i\e ]} \br
\label{LHS}
\ee
\be 
RHS \equiv
{\lambda ^2\over 4\pi} cos ^2 { \otp_1  p_2\over 2}
\int d^3 q_1 \int d^3 q_2 \, 
\delta (q^0 _2 - {{\vec {q_2}}^2 
\over {2}})  
\delta (q^0 _1 - {{\vec {q_1}}^2 \over {2}}) 
\del3 (p_1+p_2-q_1-q_2) 
 cos ^2 {\tilde{q}_1 q_2 \over 2} 
\label{preRHS}
\ee
where $\otp^{\mu} \equiv p_{\nu}\t^{\nu \mu}$,
$P^{\mu}=p_1^{\mu}+p_2^{\mu}$ and the products
are defined by $pk \equiv p^0k^0-\pvec\cdot\kvec$.
Using the identity $2cos^2x=1 + cos2x$ for the
cosinus involving integrated momenta, both
sides can be written as a sum of a planar integral
plus a non-planar one. It is straightforward to
show that the planar parts are identical in both 
sides. Therefore, the only job left is to check 
for the non-planar ones.
For the $RHS$ it gives
\be RHS|_{np}=
\frac{\lambda^2}{4}
cos^2 \otp_1p_2 \int \frac {d^3 k}{2\pi} 
\delta (P^0-k^0-\frac{(\opvec - \kvec )^2}{2}) 
\delta (k^0-\frac{\kvec^2}{2}) cos \tp k. 
\label{RHS}
\ee

\vskip 4mm

\subsection{Magnetic Case}

\indent

	Take the non-commutativity only in the two spatial
coordinates $[x,y]=i\t$. In this case we have $\tp^0=0$
and so we can take the cosinus of \bref{LHS}
out of the $k^0$ integral. Therefore, we can perform the $k^0$ integral using 
Cauchy's theorem. We are left with
\be
LHS|_{np}\,=\,
- {\lambda^2 \over 2(2\pi)^2} cos^2{\otp_1 p_2 \over 2} \,
Im \int d^2 k {cos\tp k \over P^0 - \frac{\kvec^2}{2}
- \frac{(\opvec-\kvec)^2}{2}+i\e}.
\ee

The imaginary part is extracted by using
that $(x+i\epsilon)^{-1}= 
{\mathcal{P}} \frac {1}{x}-i\pi \delta (x)$, and it is
then straightforward to show that we obtain exactly \bref{RHS}. 
It can be easily seen that these last two steps are equivalent to
replacing the internal propagators by delta functions. 
Indeed, this is nothing but a proof that
the cutting rules are valid for the magnetic case.
Notice that we have been able to check the Optical Theorem
to all orders in $\t$.

\vskip 6mm

\subsection{Electric Case}

\indent

	Now, take non-commutativity between space and time,
\ie $[t,x]=i\t$. The main difference with respect to the
magnetic case is that now $\tp^0 \neq 0$ and, therefore, 
the cosinus factor in \bref{LHS} cannot be taken out of the
$k^0$ integral.
We will find that the order zero $\t$-term is different
from the one we obtain in expanding the $RHS$ \bref{RHS}. Furthermore, 
a linear term arises, in contrast with the $RHS$, where the
first $\t$ term is quadratic.

Here one needs to go through Feynman parameters and residue integrals. 
We arrive at:

\be LHS|_{np}\, =\,
{i \lambda^2 \over 16\pi} 
\int_0 ^1 dx {1\over |1-2x|} \,\, \bl e^{i f(P,\t,x)} \CT(P^0,\t) 
+ e^{i{f(P,-\t,x)}} \CT(P^0,-\t) \br
\label{intx}
\ee
being
\be 
\label{def}
f(P,\t,x) \equiv 
\frac{|\tp_0|}{2|1-2x|} 
(2P^0(1-x)-\opvec^2x(1-x)) + 
\frac{\vec{\tp}^2}{2|\tp_0|} |1-2x| 
-\vec{\tp} \cdot \vec{P} (1-x) )
\ee
\be
\CT(P^0,\t) \equiv \Theta(\tp^0) \Theta(x-\undos) 
+ \Theta( \tp^0)\Theta(\undos-x).
\ee
where we have chosen the symbol $\Theta(x)$ to name the
step function, not to be confused with the
non-commutative parameter $\t$.

The integral \bref{intx} cannot be solved exactly. 
Anyway, every term in
\bref{def} is linear in the non-commutativity
parameter $\t$, since $\tp^0 = \t P^1$ and
$\vec{\tp}=(\t P^0 ,0)$. Therefore,
we can expand the expontentials of \bref{intx} 
in order to obtain a power series in $\t$ of the $LHS$.
However, due to the
singular behaviour of \bref{def} about $x=\undos$, one
must expand only 
the exponential of the non-singular
terms. Taking all this into account we finally obtain
\be
LHS|_{np} \, =\, \,
{\lambda^2 \over 16} \,\,
\,\,
+ \,|\t| \,{\lambda^2 \over 32\pi} \, \left(
{|P^1|\; \opvec\;^2} + {2(P^0)^2 \over  |P^1|} \right)
+ \, \, ...
\label{result}
\ee
	The first term arises from expanding a gamma
function with imaginary argument, in contrast with
the logarithms one finds in the relativistic case 
\cite{jaume1}. Its value is exactly half of its
$RHS$ counterpart \bref{RHS}, and so unitarity
is violated. The linear term is not present in
\bref{RHS} either. The non-analicity in $P^1$
should not come as a surprise, 
it is associated to the new infinite
uncertainty relations introduced by the non-commutativity.
Notice too that only the absolute value of $\t$ appears in
\bref{result}, in agreement with the original symmetry $\t
\rightarrow -\t$ in \bref{LHS}.

\section{Two Points Function and Possible New States}

\indent

Once the non-unitarity of theory has been shown, we
could try to fix it perturbatively by introducing
some new states in the theory with the right couplings.
In \cite{luis} it was shown that in the relativistic case
one could introduce the so called $\chi$-particles
to fix the unitarity failure. A similar analysis will show that there does not 
seem to be such a procedure for the non-relativistic case.

	The optical theorem for the one-to-one scattering
amplitude states that
\be \epsfig {file=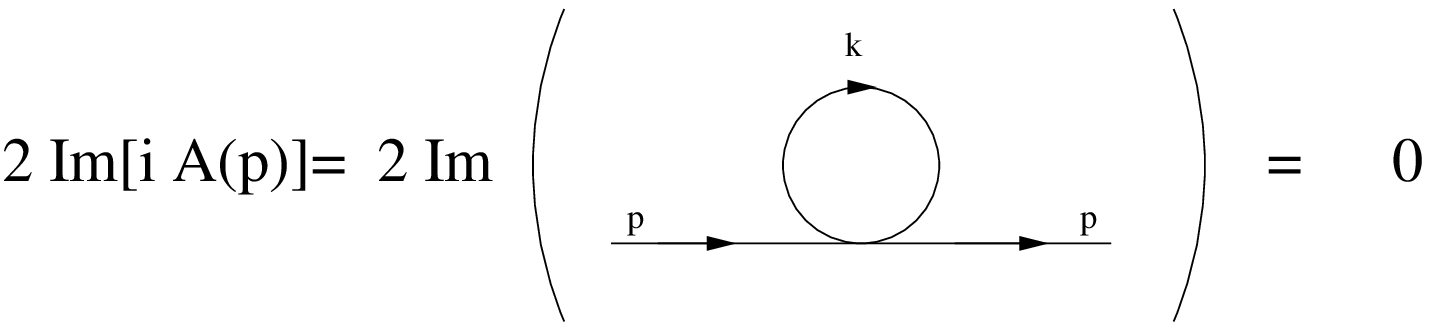, height=2cm, width=9cm} \ee

A  short calculation shows that
\be
A(p) =
-i \lambda \int {dk^0 d^2k \over (2\pi)^3}
{cos^2 {\otp k \over 2} \over k^0 - \frac{\kvec^2}{2}
+i\e} = 
{-i \lam \over 16 \pi} \, \Lambda^2 \, \, + \, \,
i {\lambda \over 8\pi}
{exp\left({ \vec{\otp}^2 \over 2 \otp^0}\right) \over |\otp^0|}.
\label{two}
\ee

In obtaining this result, we have introduced a hard cutoff
for the planar integral (it diverges as in any 
commutative theory). 
For magnetic cases, we have $\otp^0=0$. If we take this limit
in our result \bref{two}, we recover the result of 
\cite{esperanza}, \ie $A(p) = \lambda 
\delta^{(2)}(\pvec) / 4\t^2$. It has no imaginary part,
and so unitarity is preserved. On the contrary,
for electric cases $\otp^0$ is finite, and
there is always an imaginary contribution
\be
2\, Im \,[A(p)] \,=\, {\lam \over 4 \pi} \, 
{cos {\vec{\otp}^2 \over 2 \otp^0} \over |\otp^0|}.
\label{cos}
\ee

	The analog of this unitarity breakdown
in the relativistic case \cite{luis} was a delta function.
There, the authors reexpressed it as 
\be
2\, Im \,[A(p)]_{relativistic} \,=\, \int
{d^3k \over 2 (2\pi)^3 |k_1|} \rho(\lam,\t) \delta^{(4)}(p-k)
\ee
and so they interpreted it as a term
coming from the contribution of new states in the theory, 
with coupling $\rho(\lam,\t)$ to our $\phi$ field. 
It can be seen that this
cannot be done in our case even if we allow for
the coupling to depend on the momenta. 
This is due to the fact that our result \bref{cos} 
is a smooth function of the momenta, and so it
can never be written as a delta function times
a coupling.

\section{Conclusions}

\indent

	We have shown that scalar non-relativistic
non-commutative theories suffer from the
same unitarity problems as the relativistic ones. 
Magnetic cases have passed the test to all orders 
in $\theta$, while
electric ones failed in both the
two and the four points functions. The only 
relativistic theories in which non-locality in time
was compatible with unitary evolution were some
light-like $\t^{\mu\nu}$ backgrounds \cite{jaime}.
Since there are no such configurations
for non-relativistic theories (light-cone coordinates
are nonsense), non-locality in time always
destroys unitarity.

	We have also shown that
the attempt of introducing new particles to our Lagrangian
will never solve unitarity problems, not even at the
level of the one-to-one scattering. The interpretation
of the $\chi$-particles as playing the role of some
string modes that we have been missing in our original
effective field theory fails in our non-relativistic
model, which is nothing but a signal that this model cannot be
 embedded in string theory.

\vskip 6mm

{\it{\bf Acknowledgments}}

We would like to thank Joaquim Gomis for
so many fruitful discussions. This work is partially supported by
AEN98-0431 and GC 2000SGR-00026 (CIRIT). 
T.M. is supported by a fellowship from the Commissionat per 
a la Recerca de la Generalitat de Catalunya and A.M acknowledges financial 
support from a MEC FPI fellowship.

\vskip 4mm


\end{document}